\documentclass{elsart}

\usepackage{amssymb}
\usepackage{amsmath}
\usepackage{amscd}

\begin{document}

\begin{frontmatter}



\title{A formula for the Bloch vector of some Lindblad quantum systems}


\author[add]{D. Salgado\corauthref{DSF}}
\ead{david.salgado@uam.es}
\author[add]{J.L. S\'{a}nchez-G\'{o}mez}
\ead{jl.sanchezgomez@uam.es}
\address[add]{Dpto.\ F\'{\i}sica Te\'{o}rica, Univ.\ Aut\'{o}noma de Madrid\\
28049 Madrid (Spain)} 

\corauth[DSF]{Corresponding author}

\begin{abstract}
Using the Bloch representation of an $N$-dimensional quantum system and immediate results from quantum stochastic calculus, we establish a closed formula for the Bloch vector, hence also for the density operator, of a quantum system following a Lindblad evolution with selfadjoint Lindblad operators.
\end{abstract}

\end{frontmatter}

\section{Introduction}
\label{Int}
Decoherence is undoubtedly one of the main obstacles to use coherent superpositions of quantum systems as a powerful physical resource, especially in the promising quantum processing of information (cf.\ e.g.\ \cite{ChuNie00a}). Nowadays we may resort to different techniques to deal with this phenomenon \cite{Dav76a,AliLen87a,BrePet02a}, each one posing different pros and cons. In this realm, one of the most outstanding results is Lindblad-Kossakowski's theorem \cite{Lin76a,GorKosSud76a}, which gives the most general form of the evolution equation for the density matrix of an arbitrary open quantum system under certain physical assumptions, namely\footnote{Mathematical conditions such as type of convergence, etc are not included. See original references for details.} trace-preservation, complete positivity and Markovianity. The theorem states that the density operator must satisfy an evolution equation with the following structure:

\begin{subequations}
\begin{equation}\label{GenLinEqu}
\frac{d\rho(t)}{dt}=-i[H,\rho(t)]+\frac{1}{2}\left\{\sum_{k}[L_{k}\rho(t),L_{k}^{\dagger}]+[L_{k},\rho(t)L_{k}^{\dagger}]\right\}
\end{equation}

The operators $L_{k}$ are known as Lindblad operators; they are completely arbitrary and $H$ is the (Lamb-shifted \cite{BrePet02a}) Hamiltonian of the open quantum system. Though in the cited works this result was obtained from an axiomatic standpoint, other constructive approaches have also arrived at master equations with this structure \cite{IsaSanSch93a,Ali02a}.\\
In particular, the case when the Lindblad operators are Hermitian plays a prominent role in Quantum Measurement Theory \cite{BrePet02a}. In these circumstances, equation \eqref{GenLinEqu} can be rewritten as

\begin{equation}\label{HerLinEqu}
\frac{d\rho(t)}{dt}=-i[H,\rho(t)]-\frac{1}{2}\sum_{k}[L_{k},[L_{k},\rho(t)]]
\end{equation}
\end{subequations}

This equation has a clear-cut physical meaning: the system on one hand follows the unitary dynamics imposed by the Hamiltonian $H$ whereas on the other hand it also obeys the non-unitary dynamics represented by the double commutators, which tend to project the system onto the eigenspaces of $L_{k}$. When $[H,L_{k}]=0$, the preceding equation represents a typical situation of quantum non-demolition measurement.\\
On the other hand, another unvaluable tool for studying finite quantum systems is their Bloch vector \cite{Kim03a,ByrKha03a}, i.e.\ any $N$-dimensional quantum system density matrix can be expressed as $\rho=\frac{1}{N}(I_{N}+r\cdot\lambda)$, where $\lambda_{j}$ ($j=1,\dots,N^{2}-1$) denote the traceless orthogonal generators\footnote{Those elements satistying (i) $\lambda_{k}=\lambda_{k}^{\dagger}$, (ii) $\textrm{Tr}\lambda_{k}=0$ and (iii) $\textrm{Tr}\lambda_{k}\lambda_{n}=2\delta_{kn}$. See \cite{Kim03a} and references therein for details.} of $SU(N)$. This Bloch vector $r$ is restricted to satisfy certain conditions \cite{Kim03a,ByrKha03a}.\\
We dedicate this Letter to provide a closed formula for the Bloch vector $r(t)$ at an arbitrary time $t$  of an $N$-dimensional quantum system satisfying equation \eqref{HerLinEqu}. Our result rests, apart from the Bloch representation, upon the well-known fact that any evolution given by \eqref{HerLinEqu} can be understood as an averaged random unitary evolution (cf.\ e.g.\ \cite{BucHor02a}).\\
Before presenting the main results, we will clarify the viewpoint followed in this Letter in comparison especially with well-known and fully developed stochastic methods in Hilbert space. There exist two main ways of understanding the use of stochatic methods to study the evolution of quantum systems. Firstly, the 'quantum trajectory' method (or quantum jump approach) (cf. \cite{PleKni98a,BrePet02a} and numerous references therein) is profusely employed in Quantum Optics to simulate the evolution of a single open quantum system (normally an ion subjected to the action of a laser field) and can be understood as a adequate combination of both the standard unitary (Schr\"{o}dinger) evolution and the projective (generalized or not) measurements. The adequacy of this technique appears as a consequence of the current experimental feasibility of monitorizing the evolution of a single quantum system.\\ 
Secondly, a more fundamental use of these stochastic methods can be seen in the several models of dynamical reduction, especially designed to attack the measurement problem (\cite{Pea99a,BasGhi03a} and references therein). Now the evolution is treated as fundamental, with the consequent conceptual deviation from standard Quantum Mechanics.\\
Here we remain close to the former, but instead of pursuing a physically motivated simulation of the evolution of a quantum system, we try to find simple methods to obtain the \emph{analytical} solutions of previously known master equations. In this sense, the reader may consult \cite{SalSan03a} where a very simple method was presented to solve the phase-damping master equation for any arbitrary system. This establishes an important different with the former approach, where \emph{numerical} simulations have been fully developed (cf. e.g. \cite{Per98a}). We thus focus only on the mathematical advantages of these stochastic methods, trying to identify those unravellings (stochastic evolutions) which allow us to obtain the seeked solutions in a simple manner or at least good enough approximations to them. \\ 
The paper is organized as follows. In section  \ref{BloRepStoEvo} we briefly revise our needs of both the Bloch representation of an $N$-dimensional quantum system and of stochastic evolution in Hilbert space. In section \ref{CloFor} we use these tools to establish our result. Examples are worked out in section \ref{Exa}. Some conclusions are included in section \ref{Con}.


\section{Bloch representation and Stochastic Evolution in Hilbert Space}
\label{BloRepStoEvo} 

\subsection{Bloch representation of finite quantum systems}
\label{BloRep}
Given the density matrix $\rho$ of an arbitrary $N$-dimensional quantum system, one can always decompose $\rho$ using the traceless orthogonal generators of the Lie group $SU(N)$ and thus write 

\begin{equation}\label{DecDenOpe}
\rho=\frac{1}{N}\left(I_{N}+r\cdot\lambda\right)
\end{equation} 
\noindent with $r\in\mathbb{R}^{N^{2}-1}$. With this decomposition, we can assure that $\textrm{tr}\rho=1$ and $\rho^{\dagger}=\rho$, but some conditions must be imposed upon $r$ to ensure that $\rho\geq 0$, as it must be. These conditions were found in \cite{Kim03a} and \cite{ByrKha03a} and give rise to the so-called Bloch-representation space $B(\mathbb{R}^{N^{2}-1})$, i.e.\ those $(N^{2}-1)$-dimensional vectors valid to represent a quantum density matrix. In mathematical language this means that there exists a bijection $\phi$ from the set of density matrix $\mathcal{L}_{+,1}(\mathcal{H}_{N})$ to $B(\mathbb{R}^{N^{2}-1})$. This bijective application allows us to study the evolution $\mathcal{U}(t)$ of a quantum system by focusing on the evolution of its Bloch vector. This may be done by forcing the commutativity of the following diagram:

\[
\begin{CD}
\mathcal{L}_{+,1}(\mathcal{H}_{N})@>\mathcal{U}(t)>>\mathcal{L}_{+,1}(\mathcal{H}_{N})\\
@V \phi VV @VV\phi V\\
B(\mathbb{R}^{N^{2}-1})@>\Lambda(t)>>B(\mathbb{R}^{N^{2}-1})
\end{CD}
\]

\noindent i.e. by imposing $\mathcal{U}(t)=\phi^{-1}\circ\Lambda(t)\circ\phi$. In this sense, once the Bloch vector at an arbitrary time $r(t)=\Lambda(t)[r(0)]$ is known, we also know the corresponding evolved density matrix.

\subsection{Stochastic Evolution in Hilbert Space}
\label{StoEvo}

One of the mathematical tools to analyze the evolution of open quantum systems in noisy environments is the random evolution in Hilbert space. Though there is an impressive mathematical theory behind this approach \cite{Par92a}, we will only use the following results. An open system obeying equation \eqref{HerLinEqu} can be represented by a random state vector which evolves according to a stochatic differential equation such as

\begin{equation}
d|\phi_{t}\rangle=dU_{t}|\phi_{t}\rangle
\end{equation} 

\noindent where $U_{t}$ is a random evolution operator given by

\begin{equation}
U_{t}=\mathbb{T}\exp\left(-itH-i\sum_{k}L_{k}\mathcal{B}_{k}(t)\right)
\end{equation}

\noindent where $\mathbb{T}$ denotes the time-ordering opeator and $B_{k}(t)$ denotes the $k$th component of a multidimensional standard Brownian motion \cite{KarShr91a}. Under these conditions, the density operator $\rho(t)$ at an arbitrary time $t\geq 0$ is given by

\begin{equation}
\rho(t)=\mathbb{E}\left[U_{t}\rho(0)U^{\dagger}_{t}\right]
\end{equation}

In the following we will work out a formula for the Bloch vector $r(t)$ at any time $t\geq 0$ using these results. 

\section{A closed formula for the evolved Bloch vector}
\label{CloFor}

Notice that by dropping out the condition of unit trace and positivity, the Bloch decomposition can also be applied to any Hermitian operator, say $L$:

\begin{equation}
L=l_0\mathbb{I}_{N}+l\cdot\lambda\equiv\sum_{\mu=0}^{N^{2}-1}l_{\mu}\lambda_{\mu}
\end{equation}

\noindent where $\lambda_{0}\equiv\mathbb{I}_{N}$. Note that $l_{0}$ necessarily satisfies $l_{0}=\frac{\textrm{tr}L}{N}$ and $l\in\mathbb{R}^{N^{2}-1}$ is arbitrary (up to the specific details of the operator $L$). The generators $\lambda_{k}$ are related through the structure constants of the Lie algebra $\mathfrak{su}(N)$ \cite{Kim03a}:

\begin{equation}
[\lambda_{i},\lambda_{j}]=\sum_{k=1}^{N^{2}-1}c_{ijk}\lambda_{k} \quad\textrm{with } c_{ijk}=2 i f_{ijk}
\end{equation}

\noindent where $f_{ijk}$ denotes the structure constants of $\mathfrak{su}(N)$.\\
For future purposes we define a multiplication rule between elements of $\mathbb{R}^{N}$ given by

\begin{list}{}{
\renewcommand{\makelabel}[1]{\mbox{\textbf{#1.}}}
\settowidth{\labelwidth}{Definition.}
\addtolength{\leftmargin}{\labelsep}
}
\item[Definition]Let $\alpha,\beta\in\mathbb{R}^{N}$. The $\odot$-product of $\alpha$ and $\beta$, denoted by $\alpha\odot\beta$, is a new element of $\mathbb{R}^{N}$ whose $k$th component is given by
\begin{equation}
\left(\alpha\odot\beta\right)_{k}= i\sum_{i,j=1}^{N}c_{ijk}\alpha_{i}\beta_{j}\quad(=-2\sum_{i,j=1}^{N}f_{ijk}\alpha_{i}\beta_{j})
\end{equation}
\end{list}

It is clear that $(\mathbb{R}^{N},\odot)$ has a Lie-algebra structure. We also introduce a second multiplication rule by

\begin{list}{}{
\renewcommand{\makelabel}[1]{\mbox{\textbf{#1.}}}
\settowidth{\labelwidth}{Definition.}
\addtolength{\leftmargin}{\labelsep}
}
\item[Definition]
Let $\alpha,\beta\in\mathbb{R}^{N}$. The $\boxdot$-product of $\alpha$ and $\beta$, denoted by $\alpha\boxdot\beta$, is defined by

\begin{equation}
\alpha\boxdot\beta=\alpha\odot(\alpha\odot\beta)
\end{equation}
\end{list}
 It is also convenient to define the $\mathbb{T}$-exponential of $\odot$ and $\boxdot$.

\begin{list}{}{
\renewcommand{\makelabel}[1]{\mbox{\textbf{#1.}}}
\settowidth{\labelwidth}{Definition.}
\addtolength{\leftmargin}{\labelsep}
}
\item[Definition]
Let $\alpha,\beta\in\mathbb{R}^{N}$, with $\alpha$ possibly time-dependent $\alpha=\alpha(t)$. Then we define the deterministic $\mathbb{T}$-exponential of $\odot$ by 

\begin{equation}\label{OdotExp}
\mathbb{T} e^{\alpha(t)\odot}\beta\equiv\sum_{n=0}^{\infty}\frac{1}{n!}\int_{0}^{t} d t_{1}\cdots\int_{0}^{t} d t_{n}\mathbb{T}[(\alpha(t_{n})\odot(\alpha(t_{n-1})\odot(\dots(\alpha(t_{1})\odot\beta)]
\end{equation}

Equivalently for the $\boxdot$ product:

\begin{equation}\label{BoxdotExp}
\mathbb{T} e^{\alpha(t)\boxdot}\beta\equiv\sum_{n=0}^{\infty}\frac{1}{n!}\int_{0}^{t} d t_{1}\cdots\int_{0}^{t} d t_{n}\mathbb{T}[(\alpha(t_{n})\boxdot(\alpha(t_{n-1})\boxdot(\dots(\alpha(t_{1})\boxdot\beta)]
\end{equation}

\end{list}

To convince oneself that these operations are well-defined, one must notice that 

\begin{equation}
||\alpha\odot\beta||\leq (CN)^{1/2}||\alpha|| ||\beta||,
\end{equation}

\noindent where $C=\sup_{ijk}|c_{ijk}|$, implies that the series \eqref{OdotExp} and \eqref{BoxdotExp} are convergent. These definitions can also be extended to the case in which Ito integrals come into play.

\begin{list}{}{
\renewcommand{\makelabel}[1]{\mbox{\textbf{#1.}}}
\settowidth{\labelwidth}{Definition.}
\addtolength{\leftmargin}{\labelsep}
}
\item[Definition]
Let $\alpha,\beta\in\mathbb{R}^{N}$, with $\alpha$ possibly time-dependent $\alpha=\alpha(t)$. Then we define the stochatic $\mathbb{T}_{s}$-exponential of $\odot$ by 

\begin{equation}\label{StoOdotExp}
\mathbb{T}_{s} e^{\alpha(t)\odot}\beta\equiv\sum_{n=0}^{\infty}\frac{1}{n!}\int_{0}^{t} d\mathcal{B}_{t_{1}}\cdots\int_{0}^{t} d\mathcal{B}_{t_{n}}\mathbb{T}[(\alpha(t_{n})\odot(\alpha(t_{n-1})\odot(\dots(\alpha(t_{1})\odot\beta)]
\end{equation}

Equivalently for the $\boxdot$ product:

\begin{equation}\label{StoBoxdotExp}
\mathbb{T}_{s} e^{\alpha(t)\boxdot}\beta\equiv\sum_{n=0}^{\infty}\frac{1}{n!}\int_{0}^{t} d\mathcal{B}_{t_{1}}\cdots\int_{0}^{t} d\mathcal{B}_{t_{n}}\mathbb{T}[(\alpha(t_{n})\boxdot(\alpha(t_{n-1})\boxdot(\dots(\alpha(t_{1})\boxdot\beta)]
\end{equation}
\end{list}


We will focus on the Heisenberg representation of the density operator $\rho^{H}(t)$. We will first give the result for the case of a single Lindblad operator. Equation \eqref{HerLinEqu} then reads

\begin{equation}\label{Her1LinEqu}
\frac{ d\rho^{H}(t)}{ d t}=-\frac{1}{2}[L^{H}(t),[L^{H}(t),\rho^{H}(t)]]
\end{equation}

\noindent where $L^{H}(t)= e^{ i tH}L e^{- i tH}$ denotes the Lindblad operator in Heisenberg picture. Note that following the preceding decompositions we can write:

\begin{subequations}
\begin{eqnarray}
\label{DecHam}H&=&\sum_{\mu=0}^{N^{2}-1}h_{\mu}\lambda_{\mu}\equiv h_{0}\lambda_{0}+h\cdot\lambda\\
\label{DecLin}L&=&\sum_{\mu=0}^{N^{2}-1}l_{\mu}\lambda_{\mu}\equiv l_{0}\lambda_{0}+l\cdot\lambda
\end{eqnarray}
\end{subequations}

For immediate purposes we propose the following

\begin{list}{}{
\renewcommand{\makelabel}[1]{\mbox{\textbf{#1.}}}
\settowidth{\labelwidth}{Definition.}
\addtolength{\leftmargin}{\labelsep}
}
\item[Definition]
Let $\alpha\in\mathbb{R}^{N^{2}-1}$ and $H=\sum_{\mu=0}^{N^{2}-1}h_{\mu}\lambda_{\mu}$. Then
we define the Heisenberg transform of $\alpha$ by

\begin{equation}
\alpha^{H}(t)\equiv\mathbb{T} e^{h\odot}\alpha\quad \textrm{where } h=(h_{1},\cdots,h_{N^{2}-1})
\end{equation} 
\end{list}
Note that $h$ is time-independent, so that one can write $\alpha^{H}(t)=\exp\left(th\odot\right)\alpha$. Then we can prove the next

\begin{list}{}{
\renewcommand{\makelabel}[1]{\mbox{\textbf{#1.}}}
\settowidth{\labelwidth}{Theorem.}
\addtolength{\leftmargin}{\labelsep}
}
\item[Theorem]\label{Gen1LinThe}
Let $\rho^{H}(t)$ be the solution of \eqref{Her1LinEqu} with decomposition $\rho^{H}(t)=\frac{1}{N}\left(\mathbb{I}_{N}+r^{H}(t)\cdot\lambda\right)$. Then $r^{H}(t)$ is given by
\begin{equation}\label{MainRes}
r^{H}(t)=\mathbb{E}[\mathbb{T}_{s}e^{-l^{H}(t)\odot}r^{H}(0)]=\mathbb{T} e^{\frac{1}{2}l^{H}(t)\boxdot}r^{H}(0)
\end{equation}

\end{list}

\begin{list}{}{
\renewcommand{\makelabel}[1]{\mbox{\textbf{#1.}}}
\settowidth{\labelwidth}{Proof.}
\addtolength{\leftmargin}{\labelsep}
}
\item[Proof]
Using appendix \ref{ComExpVal} we can rewrite \eqref{Her1LinEqu} as 
\begin{equation}\label{Proo1}
\rho^{H}(t)=\mathbb{E}[\mathbb{T} e^{- i\int_{0}^{t}[L^{H}(s),\cdot] d\mathcal{B}_{s}}][\rho(0)]
\end{equation}

First apply decompositions \eqref{DecHam} and \eqref{DecLin} to calculate $L^{H}(t)$:

\begin{eqnarray}
L^{H}(t)&=& e^{ i t[H,\cdot]}[L]=\nonumber\\
&=&l_{0}+ e^{ i t[h\cdot\lambda,\cdot]}[l\cdot\lambda]=\nonumber\\
&=&l_{0}+(\mathbb{T} e^{h\odot}l)\cdot\lambda\equiv l_{0}+l^{H}(t)\cdot\lambda
\end{eqnarray}

Now we substitute last relation and \eqref{DecDenOpe} in \eqref{Proo1} to write

\begin{equation}
\rho^{H}(t)=\mathbb{E}\left[\mathbb{T} e^{- i l^{H}_{t}\cdot[\lambda,\cdot]}\right]\left[\frac{1}{N}\left(\mathbb{I}_{N}+r^{H}(0)\cdot\lambda\right)\right]
\end{equation}

\noindent where for brevity we have denoted $l_{t}^{H}\equiv\int_{0}^{t}l^{H}(s) d\mathcal{B}_{s}$. Now the linearity of the operator and the stochastic expectaction value yields

\begin{equation}
\rho^{H}(t)=\frac{1}{N}\left(\mathbb{I}_{N}+\mathbb{E}[\mathbb{T}_{s} e^{-l^{H}(t)\odot}r^{H}(0)]\cdot\lambda\right)
\end{equation}

\noindent which proves the desired result. The last equality in \eqref{MainRes} arises as a matter of standard computation of the Gaussian expectation value.
\end{list}

A straightforward generalization to nonMarkovian evolutions of this result can be immediately obtained:

\begin{list}{}{
\renewcommand{\makelabel}[1]{\mbox{\textbf{#1.}}}
\settowidth{\labelwidth}{Theorem}
\addtolength{\leftmargin}{\labelsep}
}
\item[Theorem]
Let $\rho^{H}(t)$ be the solution of the nonMarkovian master equation

\begin{equation}
\frac{ d\rho^{H}(t)}{ d t}=-\frac{\gamma^{2}(t)}{2}[L^{H}(t),[L^{H}(t),\rho^{H}(t)]]
\end{equation}

\noindent Let $\rho^{H}(t)=\frac{1}{N}\left(\mathbb{I}_{N}+r^{H}(t)\cdot\lambda\right)$ be its Bloch decomposition. Then $r^{H}(t)$ is given by

\begin{equation}\label{NonMarBloVec}
r^{H}(t)=\mathbb{E}[\mathbb{T}_{s} e^{- l^{H}(t)\odot}r^{H}(0)]=\mathbb{T} e^{\frac{1}{2}(l^{H}(t)\gamma(t))\boxdot}r^{H}(0)
\end{equation}

\end{list}

\begin{list}{}{
\renewcommand{\makelabel}[1]{\mbox{\textbf{#1.}}}
\settowidth{\labelwidth}{Proof}
\addtolength{\leftmargin}{\labelsep}
}
\item[Proof]
Elementary generalization of the preceding proof.
\end{list}

\section{Examples}
\label{Exa}

We include two simple examples of the use of formula \eqref{MainRes}. These examples are included only to illustrate its possible uses. A detailed exact analytical treatment via master equations of these particular examples can be found e.g. in \cite{BrePet02a}, chapter 3. For simplicity's sake we will focus on two-level systems, i.e.  $\lambda_{k}\in \mathfrak{su}(2)$, with the choice $f_{ijk}=\frac{1}{2}\epsilon_{ijk}$ (thus $\odot=-\times$). First we will find the Bloch vector of equation \eqref{HerLinEqu} with Hamiltonian $H=\omega_{0}\sigma_{z}$ and Lindblad operator $L=\sqrt{\gamma}\sigma_{z}$, where $\sigma_{k}$ denotes the corresponding Pauli matrix. This is an easy example, since both operators commute. The Heisenberg transform of $l=(0,0,\sqrt{\gamma})$ is $l^{H}(t)=e^{-th\times}l=l$, as expected. The $\boxdot$-product can be  represented by a matrix operation 

\begin{equation}
l\boxdot r^{H}(0)=-\gamma Mr^{H}(0) 
\end{equation} 

\noindent where $M=\left(\begin{smallmatrix}1&0&0\\
                                       0&1&0\\
                                       0&0&0\end{smallmatrix}\right)$. Then $r^{H}(t)$ can be easily calculated:

\begin{equation}
r^{H}(t)=e^{-\frac{t\gamma}{2}M}r^{H}(0)=(e^{-\frac{\gamma t}{2}}x^{H}(0),e^{-\frac{\gamma t}{2}}y^{H}(0),z^{H}(0))
\end{equation}

The loss of coherence is clear. As a second example, let us consider $H=\omega_{0}\sigma_{z}$ and $L=\sqrt{\gamma}\sigma_{x}$. Now they do not commute. The Heisenberg transform of $l=(\sqrt{\gamma},0,0)$ is calculated through the relation $l^{H}(t)=e^{tM}l$, where now $M=\left(\begin{smallmatrix}
0&\omega_{0}&0\\
-\omega_{0}&0&0\\
0&0&0
\end{smallmatrix}\right)$:

\begin{equation}
l^{H}(t)=\sqrt{\gamma}(\cos\omega_{0}t,-\sin\omega_{0}t,0)
\end{equation}

The $\boxdot$-product is implemented through the matrix multiplication $l^{H}(t)\boxdot r^{H}(0)=-\gamma N(t)r^{H}(0)$, where 

\begin{equation}
N(t)=\left(\begin{matrix}
\sin^{2}(\omega_{0}t)&\cos\omega_{0}t\sin\omega_{0}t&0\\
\cos\omega_{0}t\sin\omega_{0}t&\cos^{2}\omega_{0}t&0\\
0&0&1
\end{matrix}\right)
\end{equation}

Taking into account that the integrand $\mathbb{T}[l^{H}(t_{n})\boxdot(l^{H}(t_{n-1})\boxdot\dots l^{H}(t_{1})\boxdot r^{H}(0))]$ in \eqref{MainRes} is completely symmetric, we can write

\begin{equation}
r^{H}(t)=\sum_{n=0}^{\infty}\left(\frac{-\gamma}{2}\right)^{n}\int_{0}^{t}dt_{n}\int_{0}^{t_{n}}dt_{n-1}\dots\int_{0}^{t_{2}}dt_{1}N(t_{n})\dots N(t_{1})r^{H}(0)
\end{equation}

This formula allows us to compute $r^{H}(t)$ to any degree of approximation (this is especially easy with nowadays computer programs with symbolic computation routines). In the present case, the third component can be explicitly calculated: $z^{H}(t)=e^{-\frac{\gamma t}{2}}z^{H}(0)$. For the two first components we include the computation to second order in $\frac{\gamma}{\omega_{0}}$:
\begin{subequations}
\begin{eqnarray}
\hspace*{-5mm}x^{H}(t)&=&\left[1-\frac{\gamma}{2\omega_{0}}\frac{\omega_{0}t}{2}(1-\textrm{sinc}2\omega_{0}t)+\frac{1}{16}\left(\frac{\gamma}{2\omega_{0}}\right)^{2}(1+2\omega_{0}^{2}t^{2}-\cos2\omega_{0}t-2\omega_{0}t\sin2\omega_{0}t)\right]x^{H}(0)+\nonumber\\
&+&\left[-\frac{\gamma}{2\omega_{0}}\frac{\omega_{0}^{2}t^{2}}{2}\textrm{sinc}^{2}\omega_{0}t+\frac{1}{16}\left(\frac{\gamma}{2\omega_{0}}\right)^{2}(4\omega_{0}t-2\omega_{0}t\cos2\omega_{0}t-\sin2\omega_{0}t\right]y^{H}(0)\\
\hspace*{-5mm}y^{H}(t)&=&\left[1-\frac{\gamma}{2\omega_{0}}\frac{\omega_{0}t}{2}(1+\textrm{sinc}2\omega_{0}t)+\frac{1}{16}\left(\frac{\gamma}{2\omega_{0}}\right)^{2}(1+2\omega_{0}^{2}t^{2}-\cos2\omega_{0}t+2\omega_{0}t\sin2\omega_{0}t)\right]y^{H}(0)+\nonumber\\
&+&\left[-\frac{\gamma}{2\omega_{0}}\frac{\omega_{0}^{2}t^{2}}{2}\textrm{sinc}^{2}\omega_{0}t+\frac{1}{16}\left(\frac{\gamma}{2\omega_{0}}\right)^{2}(2\omega_{0}t\cos2\omega_{0}t-\sin2\omega_{0}t\right]x^{H}(0)
\end{eqnarray}  
\end{subequations}

\noindent where $\textrm{sinc}x\equiv\frac{\sin x}{x}$. These relations can be e.g. used to explicitly check the loss of quantum coherence. Working at first order in $\frac{\gamma}{\omega_{0}}$ and choosing $r^{H}(0)=(1,0,0)$ we find

\begin{equation}
(r^{H}(t))^{2}=1-\frac{\gamma t}{2}(1-\textrm{sinc}2\omega_{0}t)<1
\end{equation}

\noindent valid for small times, i.e. $\gamma t\ll 1$.

\section{Conclusions}
\label{Con} 

One of the benefits of formula \eqref{NonMarBloVec} is the possibility of performing a perturbation approach to the solution of equation \eqref{HerLinEqu}. If we explicitly introduce a parameter $\gamma$ measuring the strength of the Lindblad operator $L\rightsquigarrow\gamma L$, then from \eqref{NonMarBloVec} one can develop into power series of $\gamma$:

\begin{eqnarray}
r^{H}(t)&=&r^{H}(0)+\frac{\gamma}{2}\int_{0}^{t}dt_{1}l^{H}(t_{1})\boxdot r^{H}(0)+\nonumber\\
&+&\frac{1}{2!}\frac{\gamma^{2}}{2^{2}}\int_{0}^{t}dt_{1}\int_{0}^{t}dt_{2}\mathbb{T}[l^{H}(t_{1})\boxdot(l^{H}(t_{2})\boxdot r^{H}(0))]+O(\gamma^{3})=\nonumber\\
&=&r^{H}(0)+\frac{\gamma}{2}\int_{0}^{t}dt_{1}l^{H}(t_{1})\boxdot r^{H}(0)+\nonumber\\
&+&\frac{\gamma^{2}}{2^{2}}\int_{0}^{t}dt_{1}\int_{0}^{t_{1}}dt_{2}l^{H}(t_{1})\boxdot(l^{H}(t_{2})\boxdot r^{H}(0))+O(\gamma^{3})
\end{eqnarray}

Everything has thus been reduced to a matter of simple computation. Note that for the important case of $N=2$, i.e. of qubits, the abstract operations $\odot$ and $\boxdot$ are respectively the vector product $\times$ and nested vector product in $\mathbb{R}^{3}$.\\
Several open questions remain for future work (in progress). Firstly an adaptation of the preceding theorems to infinite-dimensional systems should be found. This obviously requires a generalized Bloch decomposition for such systems. Secondly an extension of these results to generic Lindblad equations is highly desirable. The nonHermiticity of Lindblad operators poses a fundamental obstacle to apply the preceding ideas, since that Hermiticity allows us to resort to \emph{linear} random operators. In the non-Hermitean case, one is obliged to use nonlinear stochastic differential equations \cite{SalSan02c}.\\
The stochastic methods have recently been proven to be of great help in computational tasks regarding some Lindblad equations \cite{SalSan03a}, thus one may rightfully hope that a further analysis of these methods will open new paths to obtaining the solutions to these master equations.
 
\section*{Acknowledgements}

Financial support from both Madrid Education Council (grant BOCAM 20/08/99 to DS) and Spanish Ministry of Science and Technology (project no.\ BFM2002-1414) is acknowledged.


\appendix
\section{Computation of $\mathbb{E}\left[\mathbb{T}\exp\left(\mathfrak{A}(t)\right)\right]$}
\label{ComExpVal}

Here we include the computation of the stochastic average of the unitary evolution operator given by $\mathbb{T}\exp\left(\mathfrak{A}(t)\right)$ with a \emph{random superoperator} $\mathfrak{A}(t)=\int_{0}^{t}\beta(s)\mathfrak{\mathbf{S}}(s)\cdot d\mathcal{\mathbf{B}}_s$, where $\beta(s)$ is a deterministic function, $\mathfrak{\mathbf{S}}(s)\equiv(\mathfrak{S}_{1}(s),\cdots,\mathfrak{S}_{N}(s))$ is an arbitrary superoperator and $\mathcal{\mathbf{B}}_s$ denotes multidimensional standard complex or real Brownian motion. By definition of the $\mathbb{T}$ product this is equivalent to find the expectation value of 

\begin{equation}
\sum_{n=0}^{\infty}\sum_{k_{1}\cdots k_{n}}^{N}\frac{1}{n!}\int_{0}^{t}d\mathcal{B}_{t_{1}}^{k_{1}}\cdots\int_{0}^{t}d\mathcal{B}_{t_{n}}^{k_{n}}\mathbb{T}\left[\beta(t_{1})\mathfrak{S}_{k_{1}}(t_{1})\beta(t_{2})\mathfrak{S}_{k_{2}}(t_{2})\cdots\beta(t_{n})\mathfrak{S}_{k_{n}}(t_{n})\right]
\end{equation}

Using well-known properties of the Brownian motion \cite{KarShr91a}, the expectation value reduces to

\begin{eqnarray}
\mathbb{E}[\mathbb{T} e^{\mathfrak{A}(t)}]&=&\sum_{n=0}^{\infty}\sum_{j_{1}\cdots j_{n}}^{N}\frac{1}{2^{n}n!}\int_{0}^{t}d t_{1}\cdots\int_{0}^{t}d t_{n}\mathbb{T}\left[\beta^{2}(t_{1})\mathfrak{S}_{j_{1}}^{2}(t_{1})\beta(t_{2})^{2}\mathfrak{S}_{j_{2}}^{2}(t_{2})\cdots\beta(t_{n})^{2}\mathfrak{S}_{j_{n}}^{2}(t_{n})\right]\equiv\nonumber\\
&\equiv&\mathbb{T}[ e^{\frac{1}{2}\int_{0}^{t}\beta^{2}(s)\mathbf{S}^{2}(s)d s}]
\end{eqnarray}


\end{document}